\documentclass[12pt,reqno]{article}
\usepackage{graphicx}
\usepackage{amsmath}
\usepackage[euler]{textgreek}
\usepackage[russian, english]{babel}
\usepackage{amsthm,amsmath,amssymb,color,multirow,graphicx}
\usepackage{inputenc}
\usepackage{wrapfig, float,array}
\usepackage{graphicx}
\graphicspath{ {./images/} }
\usepackage{epstopdf}
\usepackage{cite}
\usepackage{tikz}

\usepackage{amscd,amssymb,amsmath,amsthm}
\usepackage[arrow,matrix]{xy}
\usepackage{graphicx}
\usepackage{color}
\usepackage{cite}
\usepackage{tabularx}
\usepackage{enumitem}
\usepackage[english,russian]{babel}
\usepackage{amsthm,amsmath,amssymb,color,multirow,graphicx}
\usepackage{inputenc}
\usepackage{wrapfig, float,array}
\usepackage{blindtext}

\usepackage{hyperref}

\usepackage[russian]{babel}
\usepackage{amscd,amssymb,amsthm}
\usepackage{textgreek}
\usepackage{graphicx}
\usepackage{color}
\usepackage{cite}
\numberwithin{equation}{section} 

\begin {document}

\begin{center}
{\bf \large { Gibbs measures for SOS models with external field on a Cayley tree}}
\end{center}

\begin{center}
   M.M.Rahmatullaev, O.Sh.Karshiboev
\end{center}

\begin{abstract}
We consider a nearest-neighbor solid-on-solid (SOS) model, with several spin values $0,1,\ldots,m,$ $m\geq2,$ and nonzero external field, on a Cayley tree of degree $k$ (with $k+1$ neighbors). We are aiming to extend the results of \cite{rs} where the SOS model is studied with (mainly) three spin values and zero external field.  The SOS model can be treated as a natural generalization of the Ising model (obtained for $m=1$). We mainly assume that $m=2$ (three spin values) and study translation-invariant (TI) and splitting (S) Gibbs measures (GMs).(Splitting GMs have a particular Markov-type property specific for a tree.) For $m=2$, in the antiferromagnit (AFM) case, a TISGM is unique for all temperatures with an external field. In the ferromagnetic (FM) case, for $m=2,$ the number of TISGMs varies with the temperature and the external field: this gives an interesting example of phase transition.

Our second result gives a classification of all TISGMs of the Three-State SOS-Model on the Cayley tree of degree two with the presence of an external field. We show uniqueness in the case of antiferromagnetic interactions and existence of up to seven TISGMs in the case of ferromagnetic interactions, where the number of phases depends on the interaction strength and external field.

\end{abstract}


\section{Introduction}

One of the central problems in the theory of Gibbs measures is to describe infinite-volume (or) limiting Gibbs measures corresponding to a given Hamiltonian. The existence of such measures for a wide class of Hamiltonians was established in the ground-breaking work of Dobrushin (see \cite{7}). However, a complete analysis of the set of limiting Gibbs measures for a specific Hamiltonian is often a difficult problem. On a cubic lattice, for small values of $\beta=\frac{1}{T},$ where $T>0$ is the temperature, a Gibbs measure is unique (Refs. \cite{B},\cite{7}) which reflects a physical fact that at high temperatures there is no phase transition. The analysis for low temperatures requires specific assumptions on the form of the Hamiltonian.

In this paper we consider models with a nearest neighbour interaction on a Cayley tree. Models on a Cayley tree were discussed in \cite{B}, \cite{G}, \cite{Zach}-\cite{Pah}. A classical example of such a model is the Ising model, with two values of spin, $\pm 1.$ It was considered in \cite{7}, \cite{Zach2} and became a focus of active research in the first half of the '90s and afterwards; Models considered in this paper are generalisations of the Ising model and can be described as SOS (solid-on-solid) models with constraints. In the case of a cubic lattice they were analysed in \cite{2} where an analogue of the so-called Dinaburg-Mazel-Sinai theory was developed. Besides interesting phase transitions in these models, the attention to them is motivated by applications, in particular in the theory of communication networks (see, for example, \cite{Kelly}, \cite{Raman}). In \cite{shok} SOS model with spins 0, 1, 2, 3 on a Cayley tree of degree $k\geq1$ is analysed and translation-invariant, periodic splitting Gibbs measures for this model are constructed. In \cite{din} the SOS model with magnetic external field on $Z^d,~d\geq2$ is studied. In \cite{KRSOS} a classification of all translation-invariant splitting Gibbs measures of SOS model on the Binary tree with spin values $0,~1~\mbox{and}~2$ is given and whether these measures are extremal or non-extremal in the set of all Gibbs measures is investigated. In \cite{RH} Potts and SOS models with $q\geq2$ states on the Cayley tree of order $k\geq1$ are considered. For any values of the parameter $q$ in the Potts model and $q\leq6$ in the SOS model, sets containing all translation-invariant Gibbs measures are found. In \cite{Muh} the authors studied translation-invariant and periodic ground states for the SOS model with external fields. In \cite{Bun} the SOS model for $m=2$ on a Cayley tree of degree $k\geq3$ is considered and using translation-invariant Gibbs measures on a Cayley tree of degree two, some non-translation-invariant Gibbs measures are constructed. In \cite{Bun2} weakly periodic ground state for the SOS model with competing interactions on the Cayley tree of degree two are studied. In \cite{sayg}, \cite{mr} translation-invariant and periodic Gibbs measures for the Potts-SOS model on the Cayley tree are studied. In \cite{Bog} the $q$-state Potts model (i.e., with spin values in $\{1,\ldots,q\}$) on a Cayley tree of degree $k\geq2$ in an external (possibly random) field is considered. Recently, in \cite{potts}  a systematic review of the theory of Gibbs measures of Potts model on Cayley trees (developed since 2013) is given and discussed many applications of the Potts model to real world situations.

\section{ Set-up}

Following \cite{Bog} we start by summarizing the basic consepts for Gibbs measures on a Cayley tree, and also fix some notation.

\emph{Cayley tree.}  Let $\mathbb{T}^k$ be a (homogeneous) Cayley tree of degree $k\geq 2$, that is, an infinite connected cycle-free (undirected) regular graph with each vertex incident to $k+1$ edges. For example, $\mathbb{T}^1=\mathbb{Z}.$ Denote by $V=\{x\}$ the set of the vertices of the tree and by $E=\{\langle x,y \rangle\}$ the set of its (non-oriented) edges connecting pairs of neighbouring vertices. The natural distance $d(x,y)$ on $\mathbb{T}^k$ is defined as the number of edges on the unique path connecting vertices $x,y\in V.$ In particular, $\langle x,y \rangle \in E$ whenever $d(x,y)=1.$ A (non-empty) set $\Lambda\subset V$ is called \emph{connected} if for any $x,y\in\Lambda$ the path connecting $x$ and $y$ lies in $\Lambda.$ We denote the complement of $\Lambda$ by $\Lambda^c:=V\backslash\Lambda$ and its \emph{boundary} by $\partial\Lambda:=\{ x\in \Lambda^c:\exists y, d(x,y)=1 \},$ and we write $\overline{\Lambda}=\Lambda\bigcup\partial\Lambda.$ The subset of edges in $\Lambda$ is denoted $E_\Lambda:=\{\langle x,y \rangle \in E:x,y\in\Lambda \}.$

Fix a vertex $x_0\in V$, interpreted as the \emph{root} of the tree. We say that $y\in V$ is a \emph{direct successor} of $x\in V$ if $x$ is penultimate vertex on the unique path leading from the root $x_0$ to the vertex $y$; that is, $d(x_0,y)=d(x_0,x)+1$ and $d(x,y)=1$. The set of all direct successors of $x\in V$ is denoted by $S(x).$ It is convenient to work with the family of the radial subsets centred at $x_0$, defined for $n\in\mathbb{N}_0:=\{0 \}\bigcup\mathbb{N}$ by $$ V_n:=\{x\in V: d(x_0,x)\leq n \}, \quad W_n:=\{x\in V:d(x_0,x)=n\},$$
interpreted as the 'ball' and 'sphere', respectively, of radius $n$ centred at the root $x_0$. Clearly, $\partial V_n=W_{n+1}.$ Note that if $x\in W_n$ then $S(x)=\{ y\in W_{n+1}:d(x,y)=1 \}$. In the special case $x=x_0$ we have $S(x_0)=W_1.$ For short, we set $E_n:=E_{V_n}.$\newline

\textbf{Remark 1} Note that the sequence of balls ($V_n$) ($n\in \mathbb{N}_0$) is \emph{cofinal} (see \cite{G}, section 1.2, page 17, see also \cite{Bog}), that is, any finite subset $\Lambda\subset V$ is contained in some $V_n.$\newline

\emph{The SOS model and Gibbs measures.} In the SOS model with external field, the spin at each vertex $x\in V$ can take values in the set $\Phi:=\{0,1,...,m \}.$ Thus, the spin configuration on $V$ is a function $\sigma:V\rightarrow\Phi$ and the set of all configurations is $\Phi^V$. For a subset $\Lambda\subset V$, we denote by $\sigma_\Lambda:\Lambda\rightarrow\Phi$ the restriction of configuration $\sigma$ to $\Lambda,$ $$\sigma_\Lambda:=\sigma(x),\quad x\in\Lambda.$$
The SOS model with external field is defined by the formal Hamiltonian
\begin{equation}\label{Ham}
H(\sigma)=-J\sum_{\langle x,y \rangle \in E}|\sigma(x)-\sigma(y)|-\sum_{x\in V}\alpha_{\sigma(x),x},
\end{equation}
where $\sigma\in\Phi^V,~$$J\in \mathbb{R}$ and ${\textbf{\textalpha}}_x=(\alpha_{0,x},\alpha_{1,x},...,\alpha_{m,x})\in \mathbb{R}^{m+1}$ is the external (possible random) field.

Here, $J<0$ gives a ferromagnetic and $J>0$ an anti-ferromagnetic model.

For each finite subset $\Lambda\subset V$ ($\Lambda\neq\emptyset$) and any fixed subconfiguration $\eta\in\Phi^{\Lambda^c}$ (called the \emph{configurational boundary condition}), the \emph{Gibbs distribution} $\gamma^{\eta}_\Lambda$ is a probability measure in $\Phi^\Lambda$ defined by the formula
\begin{equation}\label{gamma}
\gamma^\eta_\Lambda(\varsigma)=\frac{1}{Z^\eta_\Lambda(\beta)}\exp \Big\{-\beta H_\Lambda(\varsigma)+\beta J\sum_{x\in\Lambda}\sum_{y\in\Lambda^c}|\varsigma(x)-\eta(y)| \Big \}, \quad \varsigma\in\Phi^\Lambda,
\end{equation}
where $\beta\in (0,\infty)$ is a parameter (having the meaning of \emph{inverse temperature}), $H_\Lambda$ is the restriction of the Hamiltonian (\ref{Ham}) to subconfigurations in $\Lambda,$
\begin{equation}\label{Ham2}
H_\Lambda(\varsigma)=-J\sum_{\langle x,y \rangle \in E_\Lambda}|\varsigma(x)-\varsigma(y)|-\sum_{x\in\Lambda}\alpha_{\varsigma(x),x}, \quad \varsigma\in\Phi^\Lambda,
\end{equation}
and $Z^\eta_\Lambda$ is the normalizing constant (often called the \emph{canonical partition function}),
$$
Z^\eta_\Lambda=\sum_{\varsigma\in\Phi^\Lambda}\exp \Big\{-\beta H_\Lambda(\varsigma)+\beta J\sum_{x\in\Lambda}\sum_{y\in\Lambda^c}|\varsigma(x)-\eta(y)| \Big\}.
$$
Due to the nearest-neighbour interaction, formula (\ref{gamma}) can be rewritten as
\begin{equation}\label{gamma2}
\gamma_\Lambda(\varsigma)=\frac{1}{Z^\eta_\Lambda}\exp \Big\{-\beta H_\Lambda(\varsigma)+\beta J\sum_{x\in\Lambda}\sum_{y\in\partial\Lambda}|\varsigma(x)-\eta(y)| \Big\}.
\end{equation}
Finally, a measure $\mu=\mu_{\beta,\alpha}$ on $\Phi^V$ is called a \emph{Gibbs measure} if, for any non-empty finite set $\Lambda\subset V$ and any $\eta\in\Phi^{\Lambda^c},$
\begin{equation}\label{mu}
\mu(\sigma_\Lambda=\varsigma|\sigma_{\Lambda^c}=\eta)\equiv\gamma^\eta_\Lambda(\varsigma), \quad \varsigma\in\Phi^V.
\end{equation}\newline
\emph{SGM construction.} It is convenient to construct Gibbs measures on the Cayley tree $\mathbb{T}^k$ using a version of Gibbs distributions on the balls ($V_n$) defined via auxiliary fields encapsulating the interaction with the exterior of the balls. More precisely, for a vector field $ V\ni x\mapsto h_x=(h_{0,x},...,h_{m,x})\in \mathbb{R}^{m+1}$ and each $n\in \mathbb{N}_0$, define a probability measure in $V_n$ by the formula
\begin{equation}\label{comp}
\mu^h_n(\sigma_n)=\frac{1}{Z_n}\exp\Big\{-\beta H_n(\sigma_n)+\beta\sum_{x\in W_n}h_{\sigma_n(x),x} \Big\}, \quad\sigma_n\in\Phi^{V_n},
\end{equation}
 where $Z_n=Z_n(\beta,h)$ is the normalizing factor and $H_n:=H_{V_n}$, that is (see (\ref{Ham2}))
\begin{equation}\label{Ham3}
H_n(\sigma_n)=-J\sum_{\langle x,y \rangle \in E_n}|\sigma_n(x)-\sigma_n(y)|-\sum_{x\in V_n}\alpha_{\sigma_n(x),x}, \quad \sigma_n\in\Phi^{V_n}.
\end{equation}
The vector field $\{h_x\}_{x\in V}$ in (\ref{comp}) is called \emph{generalized boundary condition (GBC)}.\newline
We say that the probability distributions (\ref{comp}) are compatible (and the intristic GBC $\{h(x)\}$ are \emph{permissible}) if for each $n\in\mathbb{N}_0$ the following identity holds,
\begin{equation}\label{comp3}
\sum_{\omega\in\Phi^{W_{n+1}}}\mu^h_{n+1}(\sigma\vee\omega)\equiv\mu^h_n(\sigma), \quad \sigma_n\in\Phi^{V_n},
\end{equation}
where the symbol $\vee$ stands for concatenation of subconfigurations. A criterion for permissibility of GBC is provided by Theorem 1 (see below). By Kolmogorov's extension theorem (see \cite{Shir}, chapter II, section 3, theorem 4, page 167), the compatibility condition (\ref{comp3}) ensures that there exists a unique measure $\mu^h=\mu^h_{\beta,\alpha}$ on $\Phi^V$ such that, for all $n\in\mathbb{N}_0$,
\begin{equation}\label{omega}
\mu^h(\sigma_{V_n}=\sigma_n)\equiv\mu^h_n(\sigma_n),\quad \sigma_n\in\Phi^{V_n},
\end{equation}
or more explicitly (substituting (\ref{comp})),
\begin{equation}\label{comp4}
\mu^h_n(\sigma_{V_n}=\sigma_n)=\frac{1}{Z_n}\exp\Big\{-\beta H_n(\sigma_n)+\beta\sum_{x\in W_n}h_{\sigma_n(x),x} \Big\}, \quad\sigma_n\in\Phi^{V_n}.
\end{equation}
\textbf{Definition.} Measure $\mu^h$ satisfying (\ref{omega}) is called a \emph{splitting Gibbs measure (SGM)}.\\
\textbf{Remark 2.} Note that adding a constant $c=c_x$ to all coordinates $h_{i,x}$ of the vector $\textbf{h}_{x}$ does not change the probability measure (\ref{comp}) due to the normalization $Z_n$. The same is true for the external field $\textbf{\textalpha}_{x}$ in the Hamiltonian (\ref{Ham}). Therefore, without loss of generality we can consider \emph{reduced GBC} $\tilde{\textbf{h}}_{x}$, for example defined as
$$\tilde{h}_{i,x}=h_{i,x}-h_{m,x}, \quad i=0,...,m-1.$$
The same remark also applies to the external field $\textbf{\textalpha}$ and its reduced version $\tilde{\textbf{\textalpha}}(x)$ defined by $$\tilde{\alpha}_{i,x}:=\alpha_{i,x}-\alpha_{m,x}\quad i=0,...,m-1.$$

\section{ Compatibility criterion}

In the view of Remark 2, when working with vectors and vector-valued functions and fields it will often be convenient to pass from a generic vector $\textbf{u}=(u_0,...,u_m)\in \mathbb{R}^{m+1}$ to a 'reduced vector' $\tilde{\textbf{u}}=(\tilde{u}_0,...,\tilde{u}_{m-1})\in \mathbb{R}^m$ by setting $\tilde{u}_i=u_i-u_m$ $(i=0,...,m-1).$

The following general statement describes a criterion for GBC $\{h_x\}_{x\in V}$ to guarantee compatibility of the measure $\{\mu^h_n \}_{n\in\mathbb{N}_0}.$\newline
 \textbf{Theorem 1.}\textit{ The probability distributions $\{\mu^h_n\}_{n\in\mathbb{N}_0}$ defined in (\ref{comp}) are compatible (and the underlying GBC $\{\textbf{h}_x\}_{x\in V}$ are permissible) if and only if the following vector identity holds
\begin{equation}\label{funct}
\tilde{\textbf{h}}_{x}=\tilde{\textbf{\textalpha}}_x+\sum_{y\in S(x)}\textbf{F}(\tilde{\textbf{h}}_{y};\theta),\quad x\in V,
\end{equation}
here and below \begin{equation}\label{theta}
 \theta=e^{\beta J} \end{equation} and
 \begin{equation}\label{bel}
 \begin{split}
 \tilde{\textbf{h}}_x=(\tilde{h}_{i,x},...,\tilde{h}_{m-1,x}),~
  \tilde{\textbf{\textalpha}}_x=(\tilde{\alpha}_{i,x},...,\tilde{\alpha}_{m-1,x}),\\
 \tilde{h}_{i,x}=\beta\Big(h_{i,x}-h_{m,x}\Big)+\beta \Big(\alpha_{i,x}-\alpha_{m,x}\Big),\\
 \tilde{\alpha}_{i,x}=\beta\Big(\alpha_{i,x}-\alpha_{m,x}\Big), \quad i=0,...,m-1,
 \end{split}
 \end{equation}
 the map $\textbf{F}(\textbf{u};\theta)=(F_0(\textbf{u};\theta),...,F_{m-1}(\textbf{u};\theta))$ is defined for $\textbf{u}=(u_0,...,u_{m-1})\in \mathbb{R}^m$ and $\theta>0$ by the formulas
\begin{equation}\label{F}
F_i(\textbf{u};\theta):=\ln\frac{\sum_{j=0}^{m-1}\theta^{|i-j|}e^{u_j}+\theta^{m-i}}{\sum_{j=0}^{m-1}\theta^{m-j}e^{u_j}+1}, \quad i=0,...,m-1.
\end{equation} }

\textbf{Proof.} For shorthand, denote temporarily $\textbf{\textzeta}_x:=\textbf{h}_x+\textbf{\textalpha}_x.$ Suppose that the compatibility condition (\ref{comp3}) holds. On substituting (\ref{comp}), it is easy to see that (\ref{comp3}) simplifies to
\begin{equation}\label{teng}
\prod_{x\in W_n}\prod_{y\in S(x)}\sum_{\omega(y)\in \Phi}\exp \Big\{\beta \Big( J|\sigma_n(x)-\omega(y)|+\zeta_{\omega(y),y} \Big) \Big\}=\frac{Z_{n+1}}{Z_n}\prod_{x\in W_n}\exp\{\beta h_{\sigma_n(x),x}\},
\end{equation}
for any $\sigma_n\in\Phi^{V_n}.$ Consider the equality (\ref{teng}) on configurations $\sigma^1_n$, $\sigma^2_n\in\Phi^{V_n}$ that coincide everywhere in $V_n$ except at vertex $x\in W_n$, where $\sigma_n^1(x)=i\leq m-1$ and $\sigma^2_n(x)=m.$ Taking the log-ratio of the two resulting relations, we obtain
\begin{equation}\label{sum}
\sum_{y\in S(x)}\ln\frac{\sum_{j\in\Phi}\exp\{\beta (J|i-j|+\zeta_{j,y})\}}{\sum_{j\in\Phi}\exp\{\beta (J|m-j|+\zeta_{j,y})\}}=\beta(h_{i,x}-h_{m,x})
\end{equation}
which is readily (\ref{funct}) in view of the notations (\ref{bel}) and (\ref{F}).
Conversely, again using (\ref{bel}) and (\ref{F}), equation (\ref{funct}) can be rewritten in the coordinate form as follows,
\begin{equation}\label{ax}
\prod_{y\in S(x)}\sum_{j=0}^{m}\exp\{\beta (J|i-j|+\zeta_{j,y} )\}=a(x)\exp\{\beta h_{i,x}\},\quad i=0,...,m-1,
\end{equation}

where (omitting the immaterial dependence on $\beta,$ $\textbf{h}$ and $\textbf{\textalpha}$) we denote
$$
a(x):=\exp\{\beta h_{m,x}\}\prod_{y\in S(x)}\sum_{j=0}^{m}\exp\{\beta(J|m-j|+\zeta_{j,y})\},\quad x\in V.
$$
Hence, using (\ref{ax}) and setting $A_n:=\prod_{x\in W_n}a(x),$ we get
\begin{equation}\label{rhso}
\begin{split}
&\sum_{\omega\in\Phi^{W_{n+1}}}\mu^h_{n+1}(\sigma\vee\omega)=\frac{\exp\{-\beta H_n(\sigma_n)\}}{Z_{n+1}}\prod_{y\in W_n}\prod_{y\in S(x)}\sum_{j=0}^{m}\exp\{\beta(J|\sigma_n(x)-j|+\zeta_{j,y})\}=\\
&=\frac{A_n}{Z_{n+1}}\exp \Big\{-\beta H_n(\sigma_n)+\beta\sum_{x\in W_n}h_{\sigma_{n(x),x}} \Big\}=\frac{A_nZ_n}{Z_{n+1}}\mu^h_n(\sigma_n).
\end{split}
\end{equation}
Finally, observe that
$$
\sum_{\sigma_n\in\Phi^{V_n}}\sum_{\omega\in\Phi^{W_{n+1}}}\mu^h_{n+1}(\sigma_n\vee\omega)=$$$$=\sum_{\sigma_{n+1}\in\Phi^{V_{n+1}}}\mu^h_{n+1}(\sigma_{n+1})=1
$$
whereas from the RHS of (\ref{rhso}) the same sum is given by
$$
\sum_{\sigma_n\in\Phi^{V_n}}\frac{A_nZ_n}{Z_{n+1}}\mu^h_n(\sigma_n)=\frac{A_n Z_n}{Z_{n+1}}.
$$
Hence, $A_n Z_n/Z_{n+1}=1$ and formula (\ref{rhso}) yields (\ref{comp3}), as required. This completes the proof of Theorem 1.

\textbf{Proposition 1.} Any measure $\mu$ with local distributions $\mu^{(h)}_n$ satisfying (\ref{comp}),(\ref{comp3}) is an SGM.

\textbf{Proof.} Straightforward.

\section{ Translation-Invariant SGMs.}

From Proposition 1 it follows that for any $\tilde{\textbf{h}}=\{\tilde{\textbf{h}}_x,x\in V \}$ satisfying (\ref{funct}) there exists a unique GM $\mu$ (with restriction $\mu^{(h)}_n$ as in (\ref{comp})) and vice versa.  We suppose that the number spin values $m+1$ is 3, i.e. $m=2$ and $\Phi=\{0,1,2\}.$

For $m=2$, it follows from (\ref{funct}) that
\begin{equation}\label{eq}
\left\{%
\begin{array}{ll}
    \tilde{h}_{0,x}=\tilde{\alpha}_{0,x}+\sum\limits_{y\in S(x)}\ln\frac{\exp{(\tilde{h}_{0,y})}+\theta\exp{(\tilde{h}_{1,y})}+\theta^2}{\theta^2\exp{(\tilde{h}_{0,y})}+\theta\exp{(\tilde{\tilde{h}}_{1,y})}+1}, \\[5mm]
    \tilde{h}_{1,x}=\tilde{\alpha}_{1,x}+\sum\limits_{y\in S(x)}\ln\frac{\theta\exp{(\tilde{h}_{0,y})}+\exp{(\tilde{h}_{1,y})}+\theta}{\theta^2\exp{(\tilde{h}_{0,y})}+\theta\exp{(\tilde{h}_{1,y})}+1}. \\
\end{array}%
\right.\end{equation}

It is natural to begin with translation-invariant solutions (\ref{eq}), i.e. to assume that $\tilde{\textbf{h}}_x=\tilde{\textbf{h}}\in \mathbb{R}^{2},~\forall x\in V.$ We suppose that the external field $\tilde{\textbf{\textalpha}}_x$ is also translation-invariant, i.e. $\tilde{\textbf{\textalpha}}_x=\tilde{\textbf{\textalpha}}=(\tilde{\alpha}_0,\tilde{\alpha}_1),~\forall x\in V$ where $\tilde{\alpha}_i\in \mathbb{R},~i=0,1$. For the sake of simplicity, we restrict ourselves to the case $\tilde{\alpha}_0=0.$ Set $z_0=\exp{(\tilde{h}_{0,x})},$ $z_1=\exp(\tilde{h}_{1,x})$ and $\lambda=\exp(\tilde{\alpha}_{1,x}),$ $x\in V.$  From (\ref{eq}) we have
\begin{equation}\label{bir}
z_0=\Big(\frac{z_0+\theta z_1+\theta^2}{\theta^2z_0+\theta z_1+1} \Big)^k,
\end{equation}
\begin{equation}\label{ikki}
z_1=\lambda\Big(\frac{\theta z_0+z_1+\theta}{\theta^2z_0+\theta z_1+1} \Big)^k.
\end{equation}
\textbf{Remark 3.} The case $\lambda=1$ is studied  in \cite{rs}.

Observe that $z_0=1$ satisfies the equation (\ref{bir}) independently of $k,\theta, z_1$ and $\lambda.$ Putting $z_0=1$ into (\ref{ikki}), we obtain
\begin{equation}\label{z1}
z_1=\lambda\Big(\frac{2\theta+z_1}{\theta^2+\theta z_1+1} \Big)^k.
\end{equation}
Set \begin{equation}\label{belg}
a=a(\theta):=\frac{2\theta^{k+1}}{\lambda}, \quad b=b(\theta):=\frac{1+\theta^2}{2\theta^2},\quad x=\frac{z_1}{2\theta}.
\end{equation}
From (\ref{z1}) we get:
\begin{equation}\label{fff}
ax=\left(\frac{1+x}{b+x}\right)^k.
\end{equation}

The Eq. (\ref{fff}) is well known in the theory of Markov chains on the Cayley tree (see, e.g., \cite{B}, Proposition 10.7), and it is easy to analyse the number of its positive solutions. According to Proposition 10.7 of \cite{B} we have the following

\textbf{Lemma 1.} Denote $b_0=b_0(k):=\left(\frac{k+1}{k-1}\right)^2$ and let $\nu\equiv\nu(a,b,k)$ be the number of solutions of the Eq. (\ref{fff}) with $a,b>0$ and $k\geq2.$

(i) If $b\leq b_0$ then for all $a>0$ the Eq. (\ref{fff}) has a unique solution, i.e. $\nu=1.$

(ii) For $b>b_0$ set
\begin{equation}\label{eq9}
a_{i}=a_{i}(b,\theta):=\frac{1}{x_{i}}\Big(\frac{1+x_{i}}{b+x_{i}}\Big)^k,\quad i=1,2,
\end{equation}
with $0<a_1<a_2$ where $0<x_1(b,\theta)<x_2(b,\theta)$ are the distinct solutions of the following quadratic equation
 \begin{equation}\label{eq10}
x^2+[2-(b-1)(k-1)]x+b=0,
\end{equation}
such that
\begin{equation}\label{eq11}
x_{1}=\frac{(b-1)(k-1)-2-\sqrt{D(b,k)}}{2},
\end{equation}
and
\begin{equation}\label{eq11}
x_{2}=\frac{(b-1)(k-1)-2+\sqrt{D(b,k)}}{2},
\end{equation}
where $D(b,k):=\Big(2-(b-1)(k-1)\Big)^2-4b=(b-1)(k-1)^2(b-b_0).$
Then the number of solution $\nu$ is defined as follows
\begin{equation}\label{eq14}
\nu=\left\{%
\begin{array}{lll}
    1~~\mbox{if}~~0<a<a_1~\mbox{or}~a>a_2, \\
    2~~\mbox{if}~~a=a_1~\mbox{or}~a=a_2, \\
    3~~\mbox{if}~~a_1<a<a_2. \\
   \end{array}%
\right.
\end{equation}

 It is easily seen from (\ref{belg}) that the condition $b>b_0$ implies that
\begin{equation}\label{eq15}
\theta<\theta_c\equiv\theta_c(k)=\frac{k-1}{\sqrt{k^2+6k+1}}.
\end{equation}
If $b>b_0$ then there are one, two or three solutions according as $\lambda\notin[\lambda_1^*,\lambda_2^*],~\lambda\in\{\lambda_1^*,\lambda_2^*\}$ or $\lambda\in(\lambda_1^*,\lambda_2^*),$ respectively, where
\begin{equation}\label{eq12}
\lambda_{1}^*=\lambda_{1}^*(\theta):=\frac{2\theta^{k+1}}{a_{2}}\quad\mbox{and}\quad\lambda_{2}^*=\lambda_{2}^*(\theta):=\frac{2\theta^{k+1}}{a_{1}}.
\end{equation}
By $card(GM)_{TI}$ we denote the number of translation-invariant splitting Gibbs measures (TISGMs) for the SOS model with external field.

\textbf{Theorem 2.} \textit{ Let $k\geq2.$ For the system \eqref{bir},\eqref{ikki} the following assertions hold
\begin{equation}\label{eq13}
card(GM)_{TI}\geq\left\{%
\begin{array}{lll}
    1~~\mbox{if}~\theta\geq\theta_c~\mbox{or}~\theta<\theta_c~\mbox{and}~\lambda\notin[\lambda_1^*,\lambda_2^*], \\
    2~~\mbox{if}~\theta<\theta_c~\mbox{and}~\lambda\in\{\lambda_1^*,\lambda_2^*\}, \\
    3~~\mbox{if}~\theta<\theta_c~\mbox{and}~\lambda\in(\lambda_1^*,\lambda_2^*), \\
   \end{array}%
\right.
\end{equation}
where $\theta_c$ is given in (\ref{eq15}) and $\lambda^{*}_{1,2}=\lambda^*_{1,2}(\theta)$ and defined by (\ref{eq12}).}

\textbf{Proposition 2.} \textit{If $J\geq0~(\theta\geq1),~\lambda>0$ then the system of equations (\ref{bir}),(\ref{ikki}) has a unique solution.}

\textbf{Proof.} Let $E=z_0+\theta z_1+\theta^2,~F=\theta^2 z_0+\theta z_1+1,$ then from (\ref{bir}) we have:
\begin{equation}\label{eq16}
(z_0-1)\Big[F^k+(\theta^2-1)(E^{k-1}+\ldots+F^{k-1})\Big]=0.
\end{equation}
Since $\theta\geq1(J\geq1),$ we deduce from (\ref{eq16}) that $z_0=1$ is the only solution. Then $b=\frac{1+\theta^2}{2\theta^2}\leq1<\left(\frac{k+1}{k-1}\right)^2.$ By Lemma 1, the Eq. (\ref{z1}) has a unique solution. Thus we have proved that the system (\ref{bir}), (\ref{ikki}) has a unique solution.

By Proposition 2 we have

\textbf{Corollary 1.}  If $\theta\geq1$ then there is a unique TISGM.

\section{  Full analysis of solutions of the system (\ref{bir}), (\ref{ikki}) in the case $k=2$}

 Assuming $k=2$ the two-dimensional fixed point of the system (\ref{bir}), (\ref{ikki})  for the two components of the boundary law can be written in terms of the convenient variables $x=\sqrt{z_0}$, $y=\sqrt{z_1}$ and $\sqrt{\lambda}=\zeta$ in the form
\begin{equation}\label{x}
x=\frac{x^2+\theta y^2+\theta^2}{\theta^2 x^2+\theta y^2+1},
\end{equation}
\begin{equation}\label{y}
y=\zeta\frac{\theta x^2+y^2+\theta}{\theta^2x^2+\theta y^2+1}.
\end{equation}
\textbf{Remark 4.} The case $\lambda=1$, that is $\zeta=1$, is analyzed  in \cite{KRSOS}.

From the Eq. (\ref{x}) we have $x=1$ or
\begin{equation}\label{x2}
\theta y^2=(1-\theta^2)x-\theta^2(x^2+1).
\end{equation}
\textbf{Remark 5.} Rearranging the RHS of \eqref{x2} we obtain that $\theta y^2=x(1-\theta^2(x+\frac{1}{x}+1))$. Since $x>0$ and $x+\frac{1}{x}\geq2$ we have that the equality (\ref{x2}) can hold if $\theta<\theta'_c=\frac{1}{\sqrt{3}}.$

\textbf{5.1 Case: $x=1$}

 In this case from the Eq. (\ref{y})  we get
\begin{equation}\label{kub}
\theta y^3-\zeta y^2+(\theta^2+1)y-2\zeta\theta=0.
\end{equation}
According to the Descartes' Rule of Signs, the Eq. (\ref{kub}) has not any negative root, has at least one positive root and has at most three positive roots. Denote
$$a=-\frac{\zeta}{\theta},~b=\frac{\theta^2+1}{\theta},~c=-2\zeta.$$
We calculate the discriminant of (\ref{kub}) as in \cite{k} taking into account $\zeta=\sqrt{\lambda}$
\begin{equation}\label{kdet}
\begin{split}\Delta'(\theta,\lambda):=-\Delta(\theta,\lambda)=4a^3c-a^2b^2-18abc+4b^3+27c^2=\\
=\frac{1}{\theta^4}\Big(8\theta \lambda^2+(71\theta^4-38\theta^2-1)\lambda+4\theta^7+12\theta^5+12\theta^3+4\theta\Big).
\end{split}
\end{equation}
It is known (see \cite{k}, Theorem 4.3.8) that if $\Delta'>0$ then the Eq. (\ref{kub}) has one real root and two imaginary roots. If $\Delta'=0$ then the Eq. (\ref{kub}) has three real roots, at least two of which are equal. If $\Delta'<0$ then the Eq. (\ref{kub}) has three distinct real roots. Note that $\Delta'(\cdot)$ is a quadratic function with respect to $\lambda.$ Denote
$$
A=A(\theta):=\frac{8\theta}{\theta^4},~B=B(\theta):=\frac{71\theta^4-38\theta^2-1}{\theta^4},~ C=C(\theta):=\frac{4\theta^7+12\theta^5+12\theta^3+4\theta}{\theta^4},
$$
 and $\theta_1\approx 0.7486$. Note that $A>0,~C>0$ for all $\theta>0$ and $B\geq0$ if $\theta\geq\theta_1$ and $B<0$ if $\theta<\theta_1$.  It is easy to see that if $B\geq0$ then $\Delta'>0$ for all $\lambda>0.$ Therefore, we consider only the case $B<0$, that is $\theta<\theta_1.$ We calculate the discriminant of quadratic function (\ref{kdet})
\begin{equation}\label{qd}
D=D(\theta):=B^2-4AC=\frac{(1-\theta)(1+\theta)(1-17\theta^2)^3}{\theta^8}.
\end{equation}
Denote  $\theta_2=\sqrt{\frac{1}{17}}\approx 0.2425$. It is easy to verify that $D\geq0$ if $\theta\leq\theta_2$ and $D<0$ if $\theta_2<\theta<\theta_1.$

Consider the following cases

\textbf{Case $0<\theta<\theta_2.$} In the case $D>0$ and $B<0$ then we solve (\ref{kdet}) with respect to $\lambda$
\begin{equation}\label{l1}
\lambda_1(\theta)=\frac{-71\theta^4+38\theta^2+1-\sqrt{(1-\theta)(1+\theta)(1-17\theta^2)^3}}{16\theta},
\end{equation}
\begin{equation}\label{l2}
\lambda_2(\theta)=\frac{-71\theta^4+38\theta^2+1+\sqrt{(1-\theta)(1+\theta)(1-17\theta^2)^3}}{16\theta}.
\end{equation}
Note that $\lambda_i(\theta)>0,~i=1,2.$ It is easy to see that if $\lambda\in(0,\lambda_1(\theta))\bigcup(\lambda_2(\theta),\infty)$ then $\Delta'>0,$ if $\lambda=\lambda_1(\theta)$ or $\lambda=\lambda_2(\theta)$ then $\Delta'=0,$ if $\lambda\in(\lambda_1(\theta),\lambda_2(\theta))$ then $\Delta'<0.$

\textbf{Case $\theta=\theta_2$.} In this case $D=0$ and $B<0$, then from \eqref{kdet} we have \begin{equation}\label{l3}
\tilde{\lambda}:=\lambda_1(\theta_2)=\frac{-71\theta_2^4+38\theta_2^2+1}{16\theta_2}=\frac{54\sqrt{17}}{289}\approx0.7704.
 \end{equation}
 It follows that if $\theta=\theta_2$ and $\lambda=\tilde{\lambda}$ then $\Delta'=0,$ if $\theta=\theta_2$ and $\lambda\neq \tilde{\lambda}$ then $\Delta' >0.$ Moreover, substituting $\sqrt{\tilde{\lambda}}$ and $\theta_2$ into the Eq. \eqref{kub} we obtain that the Eq. (\ref{kub}) has one three-fold solution.

\textbf{Case $\theta_2<\theta<\theta_1$.} In this case $D<0$ and $B<0$ then we get $\Delta'>0.$

Combining above results we summarise

\textbf{Lemma 2.} \textit{There exists a unique $\theta_2=\frac{1}{\sqrt{17}}$ such that
\begin{itemize}
    \item If $\theta\geq\theta_2$ or $\theta<\theta_2$ and $\lambda\notin[\lambda_1(\theta),\lambda_2(\theta))]$ then Eq. (\ref{kub}) has one positive solution $y_1>0$
    \item If $\theta<\theta_2$ and $\lambda\in\{\lambda_1(\theta),\lambda_2(\theta)\}$ then Eq. (\ref{kub}) has two positive solutions $y_2<y_1$
    \item If $\theta<\theta_2$ and $\lambda\in(\lambda_1(\theta),\lambda_2(\theta))$ then Eq. (\ref{kub}) has three positive solutions $y_3<y_2<y_1$.
    \end{itemize}
    where $\lambda_1(\theta)$ and $\lambda_2(\theta)$ are given by \eqref{l1} and \eqref{l2}, respectively.}

\textbf{Remark 6.} Note that Lemma 2 is a particular case of Theorem 2.

\textbf{5.2 Case: \eqref{x2} is satisfied}

By Remark 5 we should only consider the case $\theta<\theta'_c=\frac{1}{\sqrt{3}}.$ The Eq. (\ref{y}) can be written as
\begin{equation}\label{ykv}
y^2=\zeta^2\Big(\frac{\theta x^2+y^2+\theta}{\theta^2x^2+\theta y^2+1}\Big)^2.
\end{equation}
In the case when the equality (\ref{x2}) is satisfied then from the Eq. (\ref{ykv}) we get
$$
((1-\theta^2)x-\theta^2(x^2+1))\theta=\zeta^2 \Big(\frac{x}{x+1}\Big)^2
$$
which is equivalent to
\begin{equation}\label{x4}
\theta^3 x^4+\theta(3\theta^2-1)x^3+(4\theta^3+\zeta^2-2\theta)x^2+\theta(3\theta^2-1)x+\theta^3=0.
\end{equation}
Denoting $\xi=x+1/x$ from (\ref{x4}) we have
\begin{equation}\label{xi}
\theta^3\xi^2+\theta(3\theta^2-1)\xi+2\theta^3-2\theta+\zeta^2=0.
\end{equation}
Note that the Eq. (\ref{xi}) is a quadratic equation with respect to $\xi.$ We calculate its discriminant taking into account that $\zeta=\sqrt{\lambda}$
\begin{equation}\label{der}
D'=D'(\theta,\lambda):=\theta^2(\theta^4+2\theta^2+1-4\theta\lambda).
\end{equation}
The Eq. (\ref{xi}) has no solution if $D'<0$; it has a unique solution if $D'=0$ and two solutions if $D'>0.$
We note that $D'=0$ has a unique solution with respect to $\lambda$
\begin{equation}\label{lambda}
\lambda_3(\theta):=\frac{\theta^4+2\theta^2+1}{4\theta}.
\end{equation}
Thus we have the following
\begin{itemize}
\item if $\lambda<\lambda_3(\theta)$ then the Eq. (\ref{xi}) has two solutions $\xi_1<\xi_2$ with
\begin{equation}\label{xi1}
\xi_1=\xi_1(\theta,\lambda):=\frac{1-3\theta^2-\sqrt{\theta^4+2\theta^2+1-4\theta\lambda}}{2\theta^2},
\end{equation}
\begin{equation}\label{xi2}
\xi_2=\xi_2(\theta,\lambda):=\frac{1-3\theta^2+\sqrt{\theta^4+2\theta^2+1-4\theta\lambda}}{2\theta^2};
\end{equation}

\item if $\lambda=\lambda_3(\theta)$ then the Eq. (\ref{xi}) has a unique solution $\xi_1=\xi_2=\frac{1-3\theta^2}{2\theta^2};$

\item if $\lambda>\lambda_3(\theta)$ then the Eq. (\ref{xi}) has no solution.
\end{itemize}

 In accordance with $\xi=x+1/x$, the condition $\xi\geq2$ must be satisfied. Let $\lambda=\lambda_3(\theta)$. In the case we should solve the following inequality
 \begin{equation}\label{sh22}
\xi_1=\xi_2=\frac{1-3\theta^2}{2\theta^2}\geq2.
\end{equation}
Denote $\theta_3=\sqrt{\frac{1}{7}}\approx0.3780$. It follows from \eqref{sh22} that if $\theta=\theta_3$ then $\xi_1=\xi_2=2$ and if $\theta<\theta_3$ then $\xi_1=\xi_2>2.$

Let $\lambda<\lambda_3(\theta).$ Then the following cases arise

 \textbf{Case $\xi_1<2,~\xi_2=2$.} In this case we should solve
 \begin{equation}\label{ten5}
\left\{%
\begin{array}{ll}
    \xi_1=\frac{1-3\theta^2-\sqrt{\theta^4+2\theta^2+1-4\theta\lambda}}{2\theta^2}<2, \\[0.5cm]
    \xi_2=\frac{1-3\theta^2+\sqrt{\theta^4+2\theta^2+1-4\theta\lambda}}{2\theta^2}=2.\\
\end{array}%
\right.\end{equation}
 Simplifying (\ref{ten5}) we have
 \begin{equation}\label{ten6}
\left\{%
\begin{array}{ll}
    \sqrt{\theta^4+2\theta^2+1-4\theta\lambda}>1-7\theta^2, \\[0.6cm]
    \sqrt{\theta^4+2\theta^2+1-4\theta\lambda}=7\theta^2-1.\\
\end{array}%
\right.\end{equation}
It follows that the second part of (\ref{ten6}) has not any solution if $\theta\leq\theta_3.$ For $\theta>\theta_3$, solving the second part of (\ref{ten6}) we get
\begin{equation}\label{l4}
\lambda_4(\theta):=4\theta(1-3\theta^2).
\end{equation}
 This solution also satisfies the first part of (\ref{ten6}). Consider the following difference
\begin{equation}\label{o1}
\lambda_3(\theta)-\lambda_4(\theta)=\frac{\theta^4+2\theta^2+1}{4\theta}-4\theta(1-3\theta^2)=\frac{(1-7\theta^2)^2}{4\theta}.
\end{equation}
It is easily verified that if $\theta=\theta_3$ then $\lambda_3(\theta)=\lambda_4(\theta)$, and if $\theta\neq\theta_3$ then $\lambda_3(\theta)>\lambda_4(\theta).$

 \textbf{Case $\xi_1=2,~\xi_2>2$.} In this case we should solve
 \begin{equation}\label{ten1}
\left\{%
\begin{array}{ll}
    \xi_1=\frac{1-3\theta^2-\sqrt{\theta^4+2\theta^2+1-4\theta\lambda}}{2\theta^2}=2 \\[0.6cm]
    \xi_2=\frac{1-3\theta^2+\sqrt{\theta^4+2\theta^2+1-4\theta\lambda}}{2\theta^2}>2\\
\end{array}%
\right.\end{equation}
 From (\ref{ten1}) we obtain
 \begin{equation}\label{ten2}
\left\{%
\begin{array}{ll}
    \sqrt{\theta^4+2\theta^2+1-4\theta\lambda}=1-7\theta^2, \\[0.4cm]
    \sqrt{\theta^4+2\theta^2+1-4\theta\lambda}>7\theta^2-1.\\
\end{array}%
\right.\end{equation}
It is obvious that the first part of (\ref{ten2}) has not any solution if $\theta\geq\theta_3.$ For $\theta<\theta_3$, solving the first part of (\ref{ten2}) we get $\lambda=\lambda_4(\theta).$ This solution also satisfies the second part of (\ref{ten2}).

\textbf{Case $\xi_1>2,~\xi_2>2.$}
Since $\xi_1<\xi_2$ it suffices to solve the following inequality
 \begin{equation}\label{ten3}
 \xi_1=\frac{1-3\theta^2-\sqrt{\theta^4+2\theta^2+1-4\theta\lambda}}{2\theta^2}>2.
 \end{equation}
After simple algebra we have
 \begin{equation}\label{ten4}
 \sqrt{\theta^4+2\theta^2+1-4\theta\lambda}<1-7\theta^2.
\end{equation}
Let  $\theta<\theta_3$, then from (\ref{ten4}) we get $\lambda>\lambda_4(\theta).$ For $\theta\geq\theta_3$ the inequality \eqref{ten4} has no solution.

\textbf{Case $\xi_1<2,~\xi_2>2.$}
In this case we should solve
 \begin{equation}\label{tenm1}
\left\{%
\begin{array}{ll}
    \xi_1=\frac{1-3\theta^2-\sqrt{\theta^4+2\theta^2+1-4\theta\lambda}}{2\theta^2}<2, \\[0.7cm]
    \xi_2=\frac{1-3\theta^2+\sqrt{\theta^4+2\theta^2+1-4\theta\lambda}}{2\theta^2}>2.\\
\end{array}%
\right.\end{equation}
Simplifying (\ref{tenm1}) we get
 \begin{equation}\label{tenm2}
\left\{%
\begin{array}{ll}
    \sqrt{\theta^4+2\theta^2+1-4\theta\lambda}>1-7\theta^2, \\[0.4cm]
    \sqrt{\theta^4+2\theta^2+1-4\theta\lambda}>7\theta^2-1.\\
\end{array}%
\right.\end{equation}
Let  $\theta<\theta_3$, then from the first inequality of (\ref{tenm2}) we get $\lambda<\lambda_4(\theta).$ For $\theta<\theta_3$ the second  inequality is true for all $\lambda<\lambda_3(\theta).$ For $\theta=\theta_3$ the second inequality of (\ref{tenm2}) holds for $\lambda<\lambda_3(\theta_3)=\frac{16\sqrt{7}}{49}\approx 0.8639.$ For $\theta>\theta_3$ the first inequality of (\ref{tenm2}) holds for all $\lambda<\lambda_3(\theta)$ and from the second inequality of (\ref{tenm2}) we have $\lambda<\lambda_4(\theta).$

If  $\xi_1\geq2$ then we can find positive solutions to the Eq. (\ref{x4}) explicitly, i.e. we have
\begin{equation}\label{eqy}
 x_4=\frac{1}{2}\Big(\xi_1-\sqrt{\xi_1^2-4}\Big), ~~x_5=\frac{1}{2}\Big(\xi_1+\sqrt{\xi_1^2-4}\Big).
\end{equation}
Similarly, if $\xi_2\geq2$ then we can find positive solutions to the Eq. (\ref{x4}) explicitly, i.e. we have
\begin{equation}\label{eqx}
  x_6=\frac{1}{2}\Big(\xi_2-\sqrt{\xi_2^2-4}\Big),~~ x_7=\frac{1}{2}\Big(\xi_2+\sqrt{\xi_2^2-4}\Big).
\end{equation}
Now to find corresponding $y$ we need the following

\textbf{Lemma 3.} \textit{For each $x\in\{x_4,x_5,x_6,x_7\}$, if $\theta<\theta'_c=\frac{1}{\sqrt{3}}$ and $\lambda\leq\lambda_3(\theta)$  then the RHS of (\ref{x2}) is positive, i.e.
\begin{equation}\label{eq7}
(1-\theta^2)x-\theta^2(x^2+1)>0.
\end{equation}
where $\lambda_3(\theta)$ is given by \eqref{lambda}.}

\emph{Proof.}
 We shall use that $x\in\{x_4,x_5,x_6,x_7\}$:
$$
(1-\theta^2)x-\theta^2(x^2+1)=x-\theta^2(x^2+x+1)=$$
$$
=x\Big(1-\theta^2(x+\frac{1}{x}+1\Big)\Big)=x(1-\theta^2(\xi_i+1)),\quad i=1,2.
$$
In case $i=1$ we have
$$
1-\theta^2(\xi_1+1)=\frac{1}{2}\Big(1+\theta^2+\sqrt{\theta^4+2\theta^2+1-4\theta\lambda}\Big)
$$
which is positive if $\theta<\theta'_c$ and $\lambda\leq\lambda_3(\theta)$

For $i=2$ we get
$$
1-\theta^2(\xi_2+1)=\frac{1}{2}\Big(1+\theta^2-\sqrt{\theta^4+2\theta^2+1-4\theta\lambda}\Big)
$$
which is positive if $\theta<\theta'_c$ and $\lambda\leq\lambda_3(\theta)$. Using this lemma we can define
\begin{equation}\label{yi}
y_i=\frac{1}{\sqrt{\theta}}\sqrt{(1-\theta^2)x_i-\theta^2(x_i^2+1)},\quad i=4,5,6,7.
\end{equation}

Combining above results we summarise

\textbf{Lemma 4.} \textit{The following assertions hold
\begin{enumerate}
  \item Let $\theta<\theta_3.$ The system (\ref{y}), (\ref{x2}) has
  \begin{enumerate}
  \item 4 positive solutions $(x_i,y_i),~i=4,5,6,7$ if $\lambda\in(\lambda_4(\theta),\lambda_3(\theta))$
  \item 3 positive solutions $(1,y_5)~(x_i,y_i),~i=6,7$ if $\lambda=\lambda_4(\theta)$
  \item 2 positive solutions $(x_6,y_6),~(x_7,y_7)$ if $\lambda=\lambda_3(\theta)$ or $\lambda\in\Big(0,\lambda_4(\theta)\Big)$
  \item no positive solution otherwise
  \end{enumerate}
  \item Let $\theta_3\leq\theta<\theta'_c.$ The system (\ref{y}), (\ref{x2}) has
  \begin{enumerate}
  \item 2 positive solutions $(x_6,y_6),~(x_7,y_7)$ if $\lambda\in(0,\lambda_4(\theta))$
  \item 1 positive solution $(1,y_6)$ if $\lambda=\lambda_4(\theta)$
  \item no positive solution otherwise
  \end{enumerate}
  \item Let $\theta\geq\theta'_c.$ The system (\ref{y}), (\ref{x2}) has not any positive solution,
  \end{enumerate}
  where $\theta_3=\frac{1}{\sqrt{7}}$, $\theta'_c=\frac{1}{\sqrt{3}},$ $x_i$ and $y_i,~i=4,5,6,7$ are given by \eqref{eqx} and \eqref{yi}, respectively.}

\textbf{Remark 7.} Let $x=1$ be a solution of \eqref{x2}. Then it is easy to verify that the equations (\ref{x2}) and (\ref{kub}) have a common solution $y$ if $\theta<\theta'_c$ and $\lambda=\lambda_4(\theta)$.

To give an overall view of the solutions of the system \eqref{x}, \eqref{y}, we need to compare the obtained external fields. The external fields $\lambda_1(\theta),~\lambda_2(\theta)$ exist as long as $\theta\leq\theta_2,$ hence we consider only the case $\theta\leq\theta_2.$ Denote $\theta_4=\sqrt{\frac{1}{19}}\approx0.2294.$ Elementary, but cumbersome calculation shows that
\begin{itemize}
\item If $0<\theta<\theta_4$ then $\lambda_1(\theta)<\lambda_4(\theta)<\lambda_2(\theta)<\lambda_3(\theta).$
\item If $\theta=\theta_4$ then $\lambda_1(\theta)<\lambda_2(\theta)=\lambda_4(\theta)<\lambda_3(\theta).$
\item If $\theta_4<\theta\leq\theta_2$ then $\lambda_1(\theta)\leq\lambda_2(\theta)<\lambda_4(\theta)<\lambda_3(\theta).$
\end{itemize}

Summarizing we get the full characterization of solutions:

\textbf{Lemma 5.} \textit{The set of solutions to the system (\ref{x}), (\ref{y}) changes under variations of the parameters $\theta$ and $\lambda$. There are $\theta_4(\approx0.2294)$, $\theta_2(\approx0.2425)$, $\theta_3(\approx0.3780)$, $\theta'_c(\approx0.5773),$  and $\lambda_j(\theta)$,  $j=\overline{1,4}$ are defined by (\ref{l1}), (\ref{l2}),($\ref{lambda}$) and (\ref{l4}), respectively. We have the following assertions
\begin{enumerate}
    \item Let $0<\theta<\theta_4$ then the system (\ref{x}), (\ref{y}) has
    \begin{enumerate}
    \item 7 solutions $(x_i,y_i)$, $i=\overline{1,7}$ if $\lambda\in(\lambda_4(\theta),\lambda_2(\theta))$
    \item 6 solutions $(1,y_1)$, $(1,y_2),$ $(x_i,y_i),$ $i=4,5,6,7$ if $\lambda=\lambda_2(\theta)$
    \item 5 solutions $(1,y_i),$ $i=1,2,3$ and $(x_6,y_6)$, $(x_7,y_7)$ if $\lambda\in(\lambda_1(\theta),\lambda_4(\theta)]$
    \item 5 solutions $(1,y_1),$ $(x_i,y_i)$ $i=4,5,6,7$ if $\lambda\in(\lambda_2(\theta),\lambda_3(\theta))$
    \item 4 solution $(1,y_1),~(1,y_2),~(x_6,y_6),~(x_7,y_7)$ if $\lambda=\lambda_1(\theta)$
    \item 3 solutions $(1,y_1),$ $(x_6,y_6),$ $(x_7,y_7)$ if $\lambda\in(0,\lambda_1(\theta))$ or $\lambda=\lambda_3(\theta)$
    \item otherwise 1 solution $(1,y_1)$ (see Fig. 1)
    \end{enumerate}
    \begin{center}
\includegraphics[width=12cm]{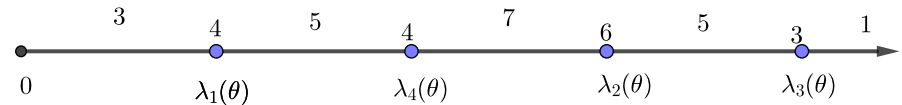}
\end{center}
\begin{center}{\footnotesize \noindent
 Figure 1. The distribution of the solutions when $\theta<\theta_4$.}\
\end{center}
    \item Let $\theta=\theta_4$ then the system (\ref{x}), (\ref{y}) has
    \begin{enumerate}
    \item 5 solutions $(1,y_i),$ $i=1,2,3$ and $(x_6,y_6),$ $(x_7,y_7)$ if $\lambda\in(\lambda_1(\theta),\lambda_2(\theta))$
    \item 5 solutions $(1,y_1),$ $(x_i,y_i),$ $i=4,5,6,7$ if $\lambda\in(\lambda_2(\theta),\lambda_3(\theta))$
    \item 4 solutions  $(1,y_1)$, $(1,y_2),$ $(x_6,y_6),$ $(x_7,y_7)$ if $\lambda=\lambda_1(\theta)$ or $\lambda=\lambda_2(\theta)$
    \item 3 solutions $(1,y_1),$ $(x_6,y_6),$ $(x_7,y_7)$ if $\lambda=\lambda_3(\theta)$ or $\lambda\in(0,\lambda_1(\theta))$
    \item otherwise 1 solution $(1,y_1)$ (see Fig. 2)
    \end{enumerate}
    \begin{center}
\includegraphics[width=12cm]{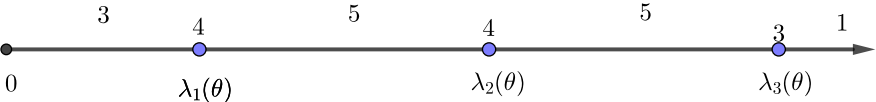}
\end{center}
\begin{center}{\footnotesize \noindent
 Figure 2. The distribution of the solutions when $\theta=\theta_4$.}\
\end{center}
    \item Let $\theta_4<\theta<\theta_2$ then the system (\ref{x}), (\ref{y}) has
    \begin{enumerate}
    \item 5 solutions $(1,y_i),$ $i=1,2,3$ and $(x_6,y_6),$ $(x_7,y_7)$ if $\lambda\in(\lambda_1(\theta),\lambda_2(\theta))$
    \item 5 solutions $(1,y_1),$ $(x_i,y_i),$ $i=4,5,6,7$ if $\lambda\in(\lambda_4(\theta),\lambda_3(\theta))$
    \item 4 solutions $(1,y_1)$, $(1,y_2),$ $(x_6,y_6),$ $(x_7,y_7)$ if $\lambda=\lambda_1(\theta)$ or
    $\lambda=\lambda_2(\theta)$
    \item  3 solutions $(1,y_1),$ $(x_6,y_6),$ $(x_7,y_7)$ if $\lambda=\lambda_3(\theta)$ or $\lambda=(0,\lambda_1(\theta))$ or $\lambda\in(\lambda_2(\theta),\lambda_4(\theta)]$
    \item otherwise 1 solution $(1,y_1)$ (see Fig. 3)
    \end{enumerate}
    \begin{center}
\includegraphics[width=12cm]{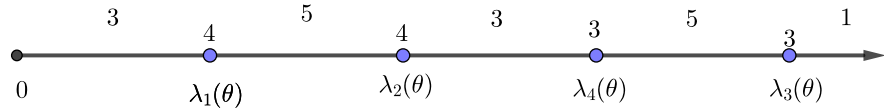}
\end{center}
\begin{center}{\footnotesize \noindent
 Figure 3. The distribution of the solutions when $\theta_4<\theta<\theta_2$.}\
\end{center}
    \item Let $\theta_2\leq\theta<\theta_3$ then the system (\ref{x}), (\ref{y}) has
    \begin{enumerate}
    \item 5 solutions $(1,y_1),$ $(x_i,y_i),$ $i=4,5,6,7$ if $\lambda\in(\lambda_4(\theta),\lambda_3(\theta))$
    \item 3 solutions $(1,y_1,)$ $(x_6,y_6),$ $(x_7,y_7)$ if $\lambda=\lambda_3(\theta)$ or $\lambda\in(0,\lambda_4(\theta)]$
    \item otherwise 1 solution $(1,y_1)$ (see Fig. 4)
    \end{enumerate}
    \begin{center}
\includegraphics[width=12cm]{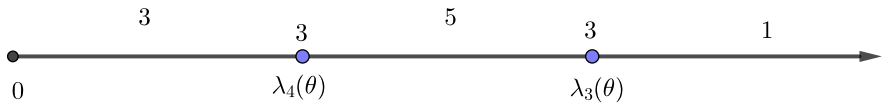}
\end{center}
\begin{center}{\footnotesize \noindent
 Figure 4. The distribution of the solutions when $\theta_2\leq\theta<\theta_4$.}\
\end{center}
    \item Let $\theta_3\leq\theta<\theta'_c$ then the system (\ref{x}), (\ref{y}) has
    \begin{enumerate}
    \item 3 solutions $(1,y_1)$, $(x_6,y_6),$ $(x_7,y_7)$ if $\lambda\in(0,\lambda_4(\theta))$
    \item otherwise 1 solution $(1,y_1)$ (see Fig. 5)
    \end{enumerate}
    \begin{center}
\includegraphics[width=12cm]{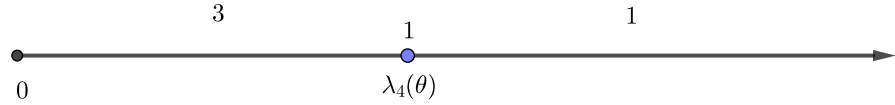}
\end{center}
\begin{center}{\footnotesize \noindent
 Figure 5. The distribution of the solutions when $\theta_3\leq\theta<\theta'_c$.}\
\end{center}
    \item Let $\theta\geq\theta'_c$ then the system (\ref{x}), (\ref{y}) has 1 solution $(1,y_1)$ for all $\lambda>0$
    \end{enumerate}
where $y_i,$ $i=1,2,3$ are solutions of the Eq. (\ref{kub}) which can be given by Cardano's formula, $x_1=x_2=x_3=1,$ $x_i$ and $y_i$ for $i=4,5,6,7$ are given by formulas (\ref{eqy}), \eqref{eqx} and (\ref{yi}).}

We denote by $\mu_i,~i=1,\ldots,7$ the TISGMs which correspond to the solutions of the system \eqref{bir}, \eqref{ikki} for $k=2.$ We have the following

\textbf{Theorem 3.} \textit{For the Three-State SOS model with external field on the Cayley tree of order two the following assertions hold: there exist $\theta_4(\approx0.2294)$, $\theta_2(\approx0.2425)$, $\theta_3(\approx0.3780)$, $\theta'_c(\approx0.5773),$  and $\lambda_j(\theta)$,  $j=\overline{1,4}$ are defined by (\ref{l1}), (\ref{l2}),($\ref{lambda}$) and (\ref{l4}), respectively, such that
\begin{enumerate}
    \item There are exactly seven TISGMs $\mu_i,~i=1,2,3,4,5,6,7$
    \begin{enumerate}
    \item if $\theta\in(0,\theta_4)$ and $\lambda\in(\lambda_4(\theta),\lambda_2(\theta))$
    \end{enumerate}
    \item There are exactly six TISGMs $\mu_i,~i=1,2,4,5,6,7$
    \begin{enumerate}
    \item if $\theta\in(0,\theta_4)$ and $\lambda=\lambda_2(\theta)$
    \end{enumerate}
    \item There are exactly five TISGMs $\mu_i,~i=1,4,5,6,7$
    \begin{enumerate}
    \item if $\theta\in(0,\theta_4]$ and $\lambda\in(\lambda_2(\theta),\lambda_3(\theta))$
    \item if $\theta\in(\theta_4,\theta_3)$ and $\lambda\in(\lambda_4(\theta),\lambda_3(\theta))$
    \end{enumerate}
    \item There are exactly another five such measures $\mu_i,~i=1,2,3,6,7$
    \begin{enumerate}
    \item if $\theta\in(0,\theta_4)$ and $\lambda\in(\lambda_1(\theta),\lambda_4(\theta)]$
    \item if $\theta\in[\theta_4,\theta_2)$ and $\lambda\in(\lambda_1(\theta),\lambda_2(\theta))$
    \end{enumerate}
    \item There are exactly four such measures $\mu_i,~i=1,2,6,7$
    \begin{enumerate}
    \item if $\theta\in(0,\theta_2)$ and $\lambda=\lambda_1(\theta)$
    \item if $\theta\in[\theta_4,\theta_2)$ and $\lambda=\lambda_2(\theta)$
    \end{enumerate}
    \item There are exactly three such measures $\mu_i,~i=1,6,7$
    \begin{enumerate}
    \item if $\theta\in(0,\theta_2)$ and $\lambda\in(0,\lambda_1(\theta))$
    \item if $\theta\in(0,\theta_3)$ and $\lambda=\lambda_3(\theta)$
    \item if $\theta\in(\theta_4,\theta_2)$ and $\lambda\in(\lambda_2(\theta),\lambda_4(\theta)]$
    \item if $\theta\in[\theta_2,\theta'_c)$ and $\lambda\in(0,\lambda_4(\theta))$
    \item if $\theta\in[\theta_2,\theta_3)$ and $\lambda=\lambda_4(\theta)$
    \end{enumerate}
    \item Otherwise there exists a unique such measure $\mu_1$
    \end{enumerate}}

\textbf{Remark 8.} If $\lambda=1$ then Theorem 3 coincides with Part 1 of the Theorem 1 in \cite{KRSOS}.

\textbf{Acknowledgments.}  The authors express their deep gratitude to Professor U. A. Rozikov for the useful advice.

\end{document}